\documentclass{article}
\usepackage{hiph-art}
\volnumber{19} \issuenumber{1} \edyear{2004}                             
\frompage{000} \topage{000}                                              
\recrevdate{30 October 2005}                                             
\def\rightmark{Transverse energy and charged particle
multiplicity}
\def\leftmark{D. Prorok}

\title{Transverse energy and charged particle
multiplicity at various centralities at RHIC: Statistical model
estimates$^{a}$}
\authors{
{Dariusz Prorok$^{1}$
\index{Prorok, D.} 
}\\[2.812mm]
{\normalsize \hspace*{-8pt}$^1$ Institute of
Theoretical Physics, University of Wroc{\l}aw,\\
PL-50-204 Wroc{\l}aw, Poland\\[0.2ex]
}}

\abstract{The transverse energy and the charged particle
multiplicity at midrapidity are evaluated in a single-freeze-out
model for different centrality bins at RHIC at $\sqrt{s_{NN}}=130$
and 200 GeV. The predictions of the model are done at the
freeze-out parameters determined earlier from measured particle
yields and $p_{T}$ spectra. The results agree qualitatively well
with the experimental data.} 


\makeindex
\begin{document}

\maketitle


The statistical model has succeeded in the description of the soft
part of particle production in heavy-ion collisions \cite{bib1}.
In particular, particle yield ratios and $p_{T}$ spectra of
identified hadrons have been reproduced with a good accuracy.
Transverse energy ($dE_{T}/d\eta$) and charged particle
multiplicity densities ($dN_{ch}/d\eta$) are global variables
whose measurements are independent of hadron spectroscopy,
therefore they could be used as an additional test of the
self-consistency of the statistical model.

The experimentally measured transverse energy is defined as

\begin{equation}
E_{T} = \sum_{i = 1}^{L} \hat{E}_{i} \cdot \sin{\theta_{i}} \;,
\label{Etdef}
\end{equation}

\noindent where $\theta_{i}$ is the polar angle, $\hat{E}_{i}$
denotes $E_{i}-m_{N}$ ($m_{N}$ means the nucleon mass) for
baryons, $E_{i}+m_{N}$ for antibaryons and the total energy
$E_{i}$ for all other particles and the sum is taken over all $L$
emitted particles \cite{bib7}.

\begin{figure}[htb]
                 \insertplot{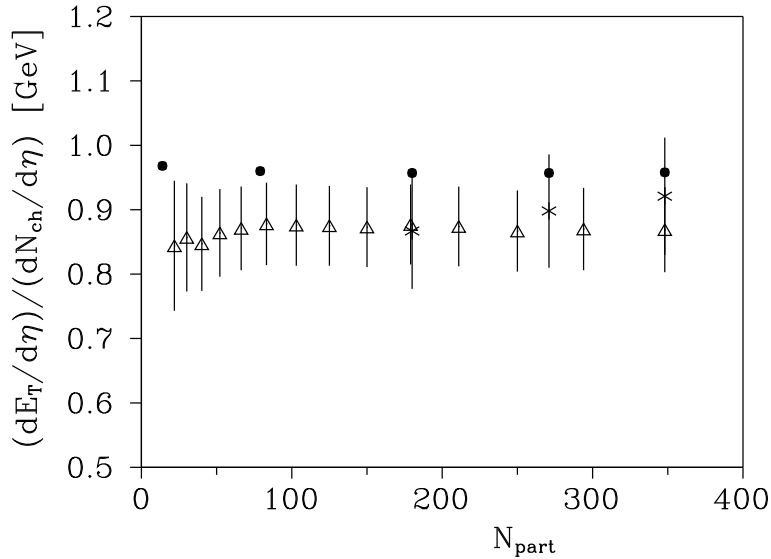}
\vspace*{-0.8cm} \caption[]{$\langle dE_{T}/d\eta\rangle /\langle
dN_{ch}/d\eta\rangle$ versus $N_{part}$ for RHIC at
$\sqrt{s_{NN}}=130$ GeV. Dots denote model evaluations, triangles
are the PHENIX data \protect\cite{bib7}. Crosses denote
recalculated PHENIX data points, \emph{i.e.} the sum of integrated
charged hadron yields \cite{bib8} have been substituted for the
denominator in the ratio. } \label{fig1}
\end{figure}

\begin{figure}[htb]
                 \insertplot{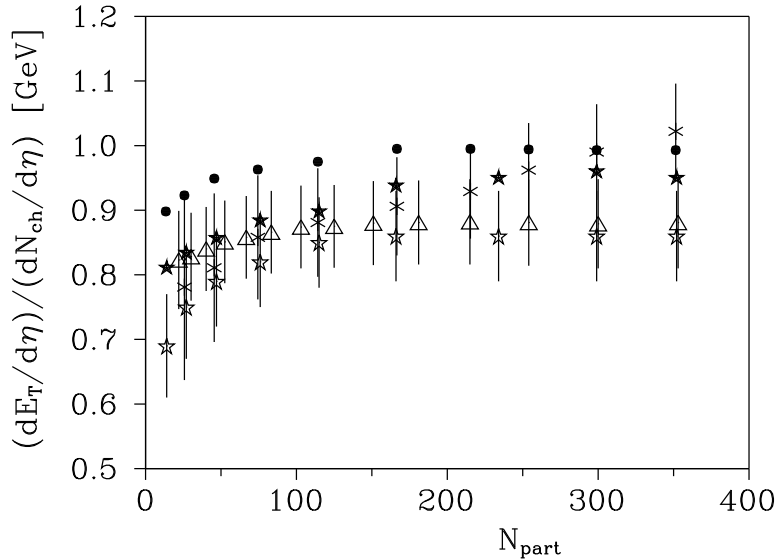}
\vspace*{-0.8cm} \caption[]{$\langle dE_{T}/d\eta\rangle /\langle
dN_{ch}/d\eta\rangle$ versus $N_{part}$ for RHIC at
$\sqrt{s_{NN}}=200$ GeV. Black dots and stars are model
evaluations for PHENIX and STAR, respectively. Triangles are the
direct PHENIX data \protect\cite{bib7}, whereas crosses are
recalculated PHENIX data points, \emph{i.e.} the sum of integrated
charged hadron yields \cite{bib9} have been substituted for the
denominator in the ratio. Open stars are the STAR data
\protect\cite{bib10}. } \label{fig2}
\end{figure}

The statistical model with single freeze-out \cite{bib2,bib3} is
applied for evaluations of $dE_{T}/d\eta$ and $dN_{ch}/d\eta$ at
midrapidity for various centrality bins at RHIC at
$\sqrt{s_{NN}}=130$ and 200 GeV. Details of this analysis can be
found elsewhere \cite{bib4,bib5}. The foundations of the model are
as follows: (\textit{a}) the chemical and thermal freeze-outs take
place simultaneously, (\textit{b}) all confirmed resonances up to
a mass of $2$ GeV from the Particle Data Tables \cite{bib6} are
taken into account, (\textit{c}) a freeze-out hypersurface is
defined by the equation $\tau =
(t^{2}-r_{x}^{2}-r_{y}^{2}-r_{z}^{2})^{1/2}= const$, (\textit{d})
the four-velocity of an element of the freeze-out hypersurface is
proportional to its coordinate, $u^{\mu}= x^{\mu} / \tau$,
(\textit{e}) the transverse size is restricted by the condition
$r=(r_{x}^{2}+r_{y}^{2})^{1/2}< \rho_{max}$. The model has four
parameters, namely, the two thermal parameters, the temperature
$T$ and the baryon number chemical potential $\mu_{B}$, and the
two geometric parameters, $\tau$ and $\rho_{max}$. Values of these
parameters were obtained from fits to particle yield ratios and
$p_{T}$ spectra (see Table 1 in \cite{bib5}, the table collects
the results from \cite{bib3,bib11}). The invariant distribution of
the measured particles of species $i$ has the Cooper-Frye form
\cite{bib2,bib3}. The distribution collects, besides the thermal
one, also contributions from simple and sequential decays such
that at least one of the final secondaries is of the \emph{i} kind
(for details, see \cite{bib3,bib4}). Having integrated this
distribution suitably over $p_{T}$ and summing up over final
particles, one can obtain $dE_{T}/d\eta$ and $dN_{ch}/d\eta$ and
finally the ratio $\langle dE_{T}/d\eta\rangle /\langle
dN_{ch}/d\eta\rangle$. The complete set of results for
$dE_{T}/d\eta$ and $dN_{ch}/d\eta$ can be found in \cite{bib5},
here only the values of the ratio as a function of the number of
participants ($N_{part}$) are shown in Figs. \ref{fig1} and
\ref{fig2}.

As one can see, the position of model predictions is very regular
and exactly resembles the configuration of the data in each case,
the estimates are only shifted up about $10\%$ as a whole. This
overestimation can be explained, at least for more central
collisions, by the observed discrepancy between the directly
measured $dN_{ch}/d\eta$ and $dN_{ch}/d\eta$ expressed as the sum
of the integrated charged hadron yields (this effect was notified
in backup slides of \cite{bib12}). If the original data points are
replaced by the recalculated data such that the denominators are
sums of the integrated charged hadron yields, then much better
agreement can be reached for more central collisions.

As far as the predictions for $dE_{T}/d\eta$ and $dN_{ch}/d\eta$
are concerned (see Figs. 1-4 in \cite{bib5}), the agreement with
the data is much better for RHIC at $\sqrt{s_{NN}}=130$ GeV. For
the case of $\sqrt{s_{NN}}=200$ GeV, only the rough qualitative
agreement has been reached. For sure, one of the reasons is that
fits in \cite{bib11} were done to the preliminary data for the
spectra \cite{bib12,bib13} and, as it turned out later, the final
data \cite{bib14,bib15} differ substantially from the preliminary
ones.

To conclude, the single-freeze-out model fairly well explains the
observed centrality dependence of transverse energy and charged
particle multiplicity pseudo-rapidity densities at midrapidity and
their ratio in the case of RHIC collisions at $\sqrt{s_{NN}}=130$
and $200$ GeV. These two variables are independent observables,
which means that they are measured independently of identified
hadron spectroscopy. It should be stressed once more that the
model fits were done earlier with the use of particle yield ratios
and $p_{T}$ spectra (not by the author, values of fitted
parameters are taken from \cite{bib3,bib11}). With the values of
parameters given, transverse energy and charged particle
multiplicity densities have been calculated in the
single-freeze-out model. Generally, the results agree
qualitatively well with the data. This adds a new argument
supporting the idea of the appearance of a thermal system during
an ultra-relativistic heavy-ion collision.

\section*{Acknowledgment}
This work was supported in part by the Polish Committee for
Scientific Research under Contract No. KBN 2 P03B 069 25.

\section*{Notes}
\begin{notes}
\item[a] This is a write-up of a poster presented at the Workshop
on Quark-Gluon-Plasma Thermalization, Vienna, Austria, 10-12
August 2005
\end{notes}

\vfill\eject
\end{document}